\documentclass[11pt,amsmath,amssymb,nofootinbib]{revtex4}
\usepackage{amsmath}
\usepackage{dcolumn}
\usepackage{bm}
\usepackage{bbm}
\usepackage[all, knot]{xy}
\xyoption{arc}

\newcommand{\sapienza}{
\affiliation{
Dipartimento di Fisica \\
 Universit\`a di Roma ``La Sapienza"\\
and Sez.~Roma1 INFN\\
P.le A. Moro 2, 00185 Roma , Italy}}

\begin{document}

\title{A no-pure-boost uncertainty principle from spacetime noncommutativity}

\author{Giovanni Amelino-Camelia}
\sapienza
\author{Giulia Gubitosi}
\sapienza
\author{Antonino\hbox{ }Marcian\`o}
\sapienza
\author{Pierre Martinetti}
\thanks{Supported by EU Marie Curie fellowship EIF-025947-QGNC}
\sapienza
\author{Flavio Mercati}
\sapienza

\begin{abstract}
\begin{center}
{\bf Abstract}
\end{center}
\noindent

\end{abstract}
\maketitle We study boost and space-rotation transformations in
$\kappa$-Minkowski noncommutative spacetime, using the techniques
that some of us had previously developed (hep-th/0607221) for a
description of translations in $\kappa$-Minkowski, which in
particular led to the introduction of translation transformation parameters
that do not commute with the spacetime coordinates. We find a
similar description  of boosts and space rotations,
which allows us to identify some associated conserved charges, but
the form of the commutators between transformation parameters and
spacetime coordinates is incompatible with the possibility of a
pure boost.
\newpage\baselineskip12pt plus .5pt minus .5pt \pagenumbering{arabic} %
\pagestyle{plain}

\section{Introduction}
A rather sizeable literature has been devoted over these past few years
to the study of the so-called $\kappa$-Minkowski noncommutative spacetime~\cite{majrue,kpoinap},
with the characteristic noncommutativity
\begin{eqnarray}
&[x_j,x_0]= i \lambda x_j \\
&[x_k,x_j]=0 ~,\label{kmnoncomm}
\end{eqnarray}
where $x_0$ is the time coordinate, $x_j$ are space coordinates ($j,k \in \{1,2,3\}$),
and $\lambda$ is a length scale, usually expected to be of the order of the
Planck length.
For most researchers involved in these studies the key source of motivation comes
from some technical observations that appear to
suggest that the symmetries of $\kappa$-Minkowski
should be described by a Hopf-algebra, the
so-called ``$\kappa$-Poincar\'e" Hopf algebra~\cite{lukieIW,majrue,kpoinap}.
However, the task of understanding the physical
implications of these Hopf-algebra $\kappa$-Poincar\'e symmetries has turned
out to be very difficult.
In particular, after more than a decade of study and hundreds of papers devoted to
the $\kappa$-Minkowski/$\kappa$-Poincar\'e framework,
even the existence of some conserved charges associated to these Hopf-algebra spacetime
symmetries was only established very recently, and only for the
translations sector
of $\kappa$-Poincar\'e,
in the analysis reported by some of us
in Ref.~\cite{k-Noether}.
This recently-developed tool of Noether analysis of some relevant Hopf-algebra symmetries,
which at least allows us to contemplate some physical observables
of the theory (the conserved charges), could of course
provide valuable elements for the debate
on the physical implications of the framework, if all of its potentialities are exploited.

With this goal in mind, we here intend to extend the analysis
reported in Ref.~\cite{k-Noether} to the full $\kappa$-Poincar\'e
Hopf algebra, thereby including also the Lorentz sector of
space-rotations and boosts. While for the description of pure
translation transformations it is necessary (as shown in
Ref.~\cite{k-Noether}) to introduce transformation parameters that
do not commute with the $\kappa$-Minkowski spacetime coordinates,
we shall show that for the case of pure space rotations it is
possible to introduce commutative transformation parameters. We
find however that the necessity of noncommutative transformation
parameters is encountered once again in the description of boost
transformations, and it takes a rather striking form: when the
boost parameters are not set to zero then also the space-rotation
parameters must not all be zero and both the boost parameters and
the space-rotation parameters must satisfy some nontrivial
commutation relations with the $\kappa$-Minkowski spacetime
coordinates. This feature could be described as a ``no-pure-boost
uncertainty principle", since it is incompatible with the
possibility of a symmetry transformation in which the only nonzero
transformation parameters are boost parameters.

This key part of our analysis is reported in the next section (Section 2).
In Section 3 we show that the transformations we introduce are genuine symmetries
of the theory, even allowing the derivation of some associated conserved charges.
In the closing Section 4 we offer a perspective on the possible implications
of our findings and consider some further studies that could take our analysis as starting point.

\section{Noncommutative transformation parameters}
\label{DifferentialCalculus}
The analysis we here report of space-rotations and boosts will be guided
by the description of translations proposed in Ref.~\cite{k-Noether}.
After the failures of several other attempts, with that description of translation transformations
it was finally possible to bring to completion a Noether analysis.
We shall therefore assume that the criteria adopted in
Ref.~\cite{k-Noether} for the description of translation transformations should
be applied also to the case of space-rotations and boosts.

In Ref.~\cite{k-Noether} the action of a translation transformation
on a function $f(x)$ of the $\kappa$-Minkowski
spacetime coordinates was parametrized as follows:
\begin{equation}
df(x)=i\epsilon^\mu P_\mu f(x) , \label{differenzialeP}
\end{equation}
 where $\epsilon_\mu$ are the transformation parameters,
 and $P_\mu$ are the Majid-Ruegg~\cite{majrue}
 translation generators, with classical action on ``time-to-the-right-ordered"
 exponentials\footnote{As conventional in the $\kappa$-Minkowski literature,
 we only give explicitly the actions on a basis of exponentials of the
 noncommutative coordinates. The action on a generic function of the $\kappa$-Minkowski
 coordinate is then induced by linearity through a Fourier-transform structure. For example,
 from (\ref{pmuaction}) one deduces~\cite{aadluna}
\begin{equation}
P_\mu f(x) = \int d^4 k \tilde{f} (k) k_\mu e^{i \vec{k} \cdot \vec{x}} e^{- i k_0 x_0}
\nonumber
\end{equation}}:
\begin{equation}
P_\mu \left( e^{i \vec{k} \cdot \vec{x}} e^{-ik_0x_0} \right)
= k_\mu e^{i \vec{k} \cdot \vec{x}} e^{-ik_0x_0} .
\label{pmuaction}
\end{equation}

The properties of the
 transformation parameters $\epsilon_\mu$
were derived~\cite{k-Noether} by
imposing Leibniz rule on the differential (\ref{differenzialeP}),
\begin{equation}
d(f(x)g(x))=(df(x))g(x)+f(x)(dg(x)) ~,
\label{leibniz}
\end{equation}
which, as a result of the observation (``coproduct"~\cite{aadluna}) that from (\ref{pmuaction})
it follows that
\begin{equation}
P_\mu \left( e^{i \vec{k} \cdot \vec{x}} e^{-ik_0x_0}
e^{i \vec{q} \cdot \vec{x}} e^{-iq_0x_0} \right)
= \left( k_\mu + e^{-\lambda k_0 (1-\delta_{\mu_0})} q_\mu \right)
\left( e^{i \vec{k} \cdot \vec{x}} e^{-ik_0x_0} e^{i \vec{q} \cdot \vec{x}} e^{-i q_0x_0} \right)
~,
\label{coprodpmu}
\end{equation}
amounts to the following requirement for the $\epsilon_\mu$:
\begin{equation}
\left( f(x) \epsilon_\mu - \epsilon_\mu e^{-\lambda P_0} f(x) \right) P_\mu g(x) = 0 ~.
\label{tleib}
\end{equation}
Clearly this equation implies that, unlike the corresponding transformation
parameters for classical Minkowski spacetime, the $\epsilon_\mu$ cannot be simply some real
numbers. Ref.~\cite{k-Noether} introduced the concept of ``noncommutative transformation
parameters" as the most conservative generalization of the concept of transformation
parameters that would allow to find solutions for (\ref{tleib}).
These noncommutative transformation parameters were required to still
act only by (associative)
multiplication on the spacetime coordinates, but were allowed
to be subject to nontrivial rules of commutation with the spacetime coordinates.
Within this generalization of the concept of transformation parameters
Eq.~(\ref{tleib}) does admit a solution, characterized by
the following rules of commutation with
the spacetime coordinates:
\begin{equation}
[\epsilon_j,x_0]=i\lambda\epsilon_j,\; [\epsilon_j,x_k]=0,\; [\epsilon_0,x_\mu]=0~.
\label{tparam}
\end{equation}

We then intend to parametrize pure space-rotation transformations in the following way:
\begin{equation}
d_R f (x) = i \sigma_j R_j f(x),
\end{equation}
where $R_j$ are the classical-action space-rotation generators~\cite{majrue,aadluna},
\begin{equation}
R_j \left( e^{i \vec{k} \cdot \vec{x}} e^{-ik_0x_0} \right)
= \epsilon_{jkl} x_k k_l e^{i \vec{k} \cdot \vec{x}} e^{-ik_0x_0}~,
\end{equation}
and the properties of the space-rotation transformation parameters $\sigma_j$
are again to be deduced from the enforcement of Leibniz rule for $d_R$.
But the well-known fact that the $R_j$ are truly classical space-rotation generators
(not only classical action but also classical ``co-action"~\cite{majrue,aadluna})
here manifests itself in the fact that the condition for enforcing
Leibniz rule for $d_R$ is trivial:
\begin{equation}
\sigma_j ( R_j f(x) ) g (x)  +  \sigma_j f(x) R_j g (x) = \sigma_j ( R_j f(x) )  g (x)
+  f(x) \sigma_j R_j g (x)~.
\end{equation}
Therefore, for pure space-rotation transformations
the $\sigma_j$ are ordinary (commutative) transformation parameters:
\begin{equation}
[\sigma_j , x_\mu] = 0~. \label{CommRotazClassici}
\end{equation}
A new level of complexity is encountered in dealing with boosts.
It is well established~\cite{majrue,aadluna} that the requirement
that the boost generators combine with the space-rotation and translation
generators to close on a Hopf algebra does not allow the introduction of
classical-action boost generators. And this requirement of a close Hopf algebra
leads to the Majid-Ruegg boost generators $N_j$:
\begin{equation}
N_j \left( e^{i \vec{k} \cdot \vec{x}} e^{-ik_0x_0} \right)
= - k_j e^{i \vec{k} \cdot \vec{x}} e^{-ik_0x_0} x_0
+ \left[ x_j \left( \frac{1 - e^{- 2 \lambda k_0}}{2 \lambda}
+ \frac{\lambda}{2} |\vec{k}|^2\right)
- \lambda x_l k_l k_j \right] e^{i \vec{k} \cdot \vec{x}} e^{-ik_0x_0} ~.
\end{equation}
We should therefore set up the description of a pure boost as follows:
\begin{equation}
d_B f(x)=i\tau_j N_j f(x),
\end{equation}
in terms of the Majid-Ruegg boost generators $N_j$
and of some transformation parameters $\tau_j$ such to enforce Leibniz rule
on $d_B$.
But the form of the Majid-Ruegg boost generators $N_j$ is such that the
Leibniz-rule requirement,
\begin{equation}
[f(x)\tau_j-\tau_j e^{-\lambda P_0}f(x)]N_j g(x)=\lambda \tau_j \epsilon_{jkl}(P_k f(x)) (R_l g(x))~,
\label{leibnogood}
\end{equation}
does not admit any acceptable solution $\tau_j$. [(\ref{leibnogood}) would require the $\tau_j$ to
be operators with highly nontrivial action on functions of the spacetime coordinates, rather
than being ``noncommutative parameters"
that act by simple (associative) multiplication on the spacetime coordinates.]

We conclude that whereas pure translations and pure space rotations are allowed
in $\kappa$-Minkowski, the possibility of pure boosts is excluded.

It is natural then to wonder whether boosts are at all allowed: one cannot have a pure boost,
but can one have transformations which combine boosts and other transformations?
We found that this is allowed and the way in which the formalism allows it is rather
intriguing. It is sufficient to contemplate a transformation that involves both a boost
and a space rotation:
\begin{equation}
d_{L}f(x)=i\tau_j N_j f(x)+i\sigma_k R_k f(x) ~,
\end{equation}
since remarkably the Leibniz-rule requirement for this ``pure-Lorentz differential",
\begin{equation}
[\tau_j (e^{-\lambda P_0}f)-f \tau_j]N_j g
+ \Big[ \lambda \epsilon_{jkl} \tau_j (P_k f) +[\sigma_l ,f]\Big] (R_l g)=0
~,
\end{equation}
does admit solutions. It is however necessary that both the $\tau_j$ and the $\sigma_j$
be ``noncommutative parameters":
 \begin{equation}
 \left\{
 \begin{array}{l}
\left[\tau_j,x_k \right] = 0 \\
\left[\tau_j,x_0 \right] = i \lambda \tau_j
   \end{array} \right.
\qquad
 \left\{
 \begin{array}{l}
 \left[ \sigma_j, x_k \right] = i \lambda \epsilon_{jlk} \tau_l \\
 \left[ \sigma_j, x_0 \right] = 0
 \end{array} \right. .
  \label{CommSigmaTauX}
 \end{equation}
From the broader perspective of these general Lorentz transformations one also
acquires a better understanding of what emerged for pure space rotations and pure boosts.
This is codified in
the commutation relations $\left[ \sigma_j, x_k \right] = i \lambda \epsilon_{jlk} \tau_l$,
which admit the case of a pure space rotation, with commutative transformation parameters
($\tau_l =0 \rightarrow \left[ \sigma_j, x_k \right] = 0$),
but are incompatible with the case of a pure boost
($\tau_l \neq 0 \rightarrow \left[ \sigma_j, x_k \right] \neq 0 \rightarrow \sigma_j \neq 0$).

\section{Noether analysis}
In order to give some substance to our claim that the transformations we constructed
in the previous section are good candidates as symmetries of theories
in $\kappa$-Minkowski spacetime, we now derive associated conserved charges for
the most studied~\cite{kpoinap,kowaorder,aadluna}
 theory formulated
in  $\kappa$-Minkowski: a theory for a massless scalar field $\Phi(x)$
governed by the Klein-Gordon-like equation of motion
\begin{equation}
 \square_\lambda \Phi (x) \equiv \tilde{P}_\mu \tilde{P}^\mu \Phi
  \equiv \left[ -\left( \frac{2}{\lambda} \right)^2 \sinh^2\left( \frac{\lambda P_0}{2}\right)
  +e^{\lambda P_0}|\vec P|^2 \right] \Phi(x) =0,\label{equescionovdemoscion}
\end{equation}
where the operator $\square_\lambda$ is the ``mass Casimir" of
the $\kappa$-Poincar\'e Hopf algebra\footnote{Eq.~(\ref{equescionovdemoscion}) reduces to the
Klein-Gordon equation in the $\lambda \rightarrow 0$ limit, and its form
was proposed (see, {\it e.g.}, Refs.~\cite{kpoinap,kowaorder,aadluna})
using as guidance the idea that it should involve an operator that commutes with all the
generators in the $\kappa$-Poincar\'e Hopf algebra.}
and we introduced the convenient notation $\tilde{P}_{\mu}$,
\begin{equation}
\tilde{P}_{0}=\left(\frac{2}{\lambda}\right)\sinh(\lambda {P}_{0}/2 ) \qquad  \tilde{P}_{j}
=e^{\lambda {P}_{0}/2} {P}_{j} ~,
\end{equation}

The equation of motion (\ref{equescionovdemoscion}) is indeed
invariant\footnote{Note that the mass Casimir ${\square_\lambda}$
commutes with all the generators
$P_\mu$,$R_j$,$N_j$ of the Hopf algebra and
with the transformation parameters $\epsilon_\mu$, $\sigma_j$, $\tau_j$.}($\delta ( {\square_\lambda} \Phi)
= {\square_\lambda} \delta \Phi = 0$)
under our transformations with $x_\mu\rightarrow x_\mu+i \epsilon^\nu P_\nu x_\mu+
i \tau_j N_j x_\mu+i\sigma_k R_k x_\mu$
and  $\Phi \rightarrow \Phi + \delta \Phi = \Phi - d_{tot}\Phi
\equiv \Phi-i\left [\epsilon^\mu P_\mu + \sigma_j R_j + \tau_k N_k \right]\Phi$
(valid when the field is a scalar under the coordinate transformation).

Our Noether analysis takes as starting point the action
\begin{equation}
S = \frac{1}{2} \int d^4x \, \Phi(x) \, {\square_\lambda}\, \Phi (x), \label{azione} \
\end{equation}
from which the equation of motion (\ref{equescionovdemoscion}) can be obtained
variationally\cite{k-Noether}.
And we focus on the Lorentz sector derived here
\begin{equation}
\delta\Phi= -d_L \Phi \equiv \left[i\sigma_j R_j + i\tau_k N_k \right] \Phi
~, \label{variescion}
\end{equation}
since for the translations one can of course follow\footnote{Since we established in
the previous section that both pure translation
transformations
and pure Lorentz-sector (space-rotation/boost)
transformations are allowed, one can indeed treat these two types of
transformations separately. But of course, they can also be analyzed simultaneously,
at the only cost of writing longer formulas.}
 the Noether analysis
reported in  detail
in Ref.~\cite{k-Noether}.

The result of a variation of the action (\ref{azione}) under our general space-rotation and boost
transformation is:
\begin{equation}
\delta S = \frac{1}{2} \int d^4 x \,  \tilde{P}^\mu \left\lbrace  \tilde{P}_\mu \left[ \left(
e^{\lambda P_0} \Phi \right) \delta \Phi \right]
- 2  \left( e^{\lambda P_0} \tilde{P}_\mu \Phi \right)
e^{\frac{\lambda}{2} P_0} \delta \Phi\right \rbrace,\label{variescionovdeacscion}
\end{equation}
where $\delta \Phi$ is given in (\ref{variescion}),
we already specialized to fields that are solutions of the equation of motion,
and we used the following property of the operators  $\tilde P_\mu$:
\begin{equation}
\tilde{P}_\mu \left[ f(x) g(x) \right] = \left[ \tilde{P}_\mu f(x)\right] \left[  e^{\frac{\lambda}{2} P_0} g(x)\right]  + \left[ e^{-\frac{\lambda}{2} P_0} f(x)\right] \left[  \tilde{P}_\mu g(x)\right]  ~.
\end{equation}
Using the fact that
from the  rules of commutation (\ref{CommSigmaTauX}) between transformation parameters and
spacetime coordinates it follows that, for a generic function of the
coordinates $f(x)$, one has $f(x) \tau_j = \tau_j (e^{-\lambda P_0}f(x))$ and $[f(x) , \sigma_j]
= \lambda\epsilon_{jlk} \tau_l (P_k f(x))$, and the observation~\cite{k-Noether}
\begin{equation}
\int d^4 x \, e^{\xi P_0} \left[ f(x) \right] = \int d^4x \,f(x) ~~~~~~~~ \forall \xi  ~,
\end{equation}
one can rewrite $\delta S$ in the following form:
\begin{equation}
\delta S =  \int d^4 x \left( i \tau_k P_\mu J^\mu_k
+ i \sigma_j P_\mu K^\mu_j \right), \label{Quadricorrenti}
\end{equation}
where:
\begin{eqnarray}
J^\mu_j (x) &=& \frac{1}{2} \left(\tilde{P}^\mu \Phi
e^{\frac{\lambda}{2} P_0}  N_j \Phi - e^{-\frac{\lambda}{2} P_0} \Phi  \tilde{P}^\mu N_j \Phi \right)
+ \nonumber
\\
&&+ \frac{\lambda}{2} \epsilon_{jkl} \left(e^{\lambda P_0} \tilde{P}^\mu P_k \Phi
e^{\frac{\lambda}{2} P_0}  R_l \Phi  - e^{\frac{\lambda}{2} P_0} P_k \Phi  \tilde{P}^\mu
R_l \Phi  \right)~,
\\
K^\mu_j (x) &=&  \frac{1}{2} \left( e^{\lambda P_0} \tilde{P}^\mu \Phi
e^{\frac{\lambda}{2} P_0}  R_j \Phi -
e^{\frac{\lambda}{2} P_0}\Phi \;  \tilde{P}^\mu R_j \Phi   \right).
\end{eqnarray}
And by spatial integration of the $J^0_j(x)$ and the $K^0_j(x)$
\begin{eqnarray}
Q^{N}_j \equiv \int d^3 x \, J^0_j(x) , \qquad Q^{R}_j \equiv \int d^3 x \, K^0_j(x) ,
\end{eqnarray}
we do obtain six time-independent (conserved) charges.  The verification of
this time independence
of the charges is most conveniently done
by writing a generic solution~\cite{k-Noether} of the
equation of motion (\ref{equescionovdemoscion}) in the form\footnote{Analogously to the
notation $\tilde{P}_\mu$ previously introduced for frequently occurring combinations of
the generators $P_\mu$, we also use the notation $\tilde{q}_0 = \frac{2}{\lambda}
\sinh \left( \frac{\lambda}{2} q_0\right)$, $\tilde{q}_j = e^{\frac{\lambda}{2} q_0} q_j$
to write more compactly some frequently occurring combinations of Fourier parameters.}:
\begin{equation}
\Phi (x) =  \int d^4 q \, \delta(\tilde{q}_\mu \tilde{q}^\mu ) \,  \tilde{\Phi} (q)  \,
e^{i \vec{q}\cdot \vec{x}} e^{- i q_0 x_0} ~.
\end{equation}
This allows us to rewrite the charges as:
\begin{eqnarray}
Q^{N}_j &=& \frac{1}{2} \int d^4k \, d^4 q \,
\delta(\tilde{k}^2) \,
 \delta(\tilde{q}^2) \; \Phi(k) \left[\lambda  \epsilon_{jlm} k_l {\mathcal{A}}_m[\tilde{\Phi}(q)]
 e^{\lambda k_0}+{\mathcal{B}}_j[\tilde{\Phi}(q)] \right] \cdot \nonumber
\\&&\qquad  \cdot \left(\tilde{k}_0  e^{\frac{\lambda}{2} q_0}
- e^{-\frac{\lambda}{2} k_0} \tilde{q}_0 \right) \delta^{(3)}(\vec{k}
 + e^{- \lambda k_0} \vec{q}) e^{- i (k_0 + q_0) x_0},
\\
Q^{R}_j &=& \frac{1}{2} \int d^4k \, d^4 q \, \delta(\tilde{k}^2)
 \, \delta(\tilde{q}^2)\; \Phi(k) {\mathcal{A}}_m[\tilde{\Phi}(q)]
   e^{\lambda k_0}  \cdot \nonumber
\\&&\qquad  \cdot \left(\tilde{k}_0  e^{\frac{\lambda}{2} q_0} -
e^{-\frac{\lambda}{2} k_0} \tilde{q}_0 \right)  \delta^{(3)}(\vec{k}
+ e^{- \lambda k_0} \vec{q}) e^{- i (k_0 + q_0) x_0},
\end{eqnarray}
where, for compactness, we introduced
\begin{equation}
{\mathcal{A}}_j[\tilde{\Phi}(q)] \equiv
i \epsilon_{jkl} q_l \frac{\partial \tilde{\Phi}(q)}{\partial q_k}
\end{equation}
and
\begin{equation}
{\mathcal{B}}_j[\tilde{\Phi}(q)] \equiv  i  q_j  \frac{\partial  \tilde{\Phi}(q)  }{\partial q_0}
+ i \left( \frac{1-e^{-2\lambda q_0}}{2 \lambda}
+ \frac{\lambda}{2} |\vec{q}|^2 \right)  \frac{\partial \tilde{\Phi}(q)}{\partial q_j}
- i \lambda q_j \left( q_k \frac{\partial \tilde{\Phi}(q)}{\partial q_k} + 2\tilde{\Phi}(q) \right) ~.
\end{equation}
With the charges $Q^{N}_j$,$Q^{R}_j$ written in this form it is then straightforward\footnote{The
details
of this derivation will be reported elsewhere~\cite{flaviotesi}.}
to verify that they indeed do not depend on the time coordinate $x_0$, and to derive
the following explicitly time-independent formulas:
\begin{eqnarray}
&Q^{N}_j = \frac{i}{2} \int   d^4 k \, \frac{\tilde{k}_0 e^{ \lambda k_0}}{|\tilde{k}_0|}
 \tilde{\Phi}(-k_0 ,-e^{\lambda k_0} \vec{k})  \left\{   k_j  \frac{\partial  \tilde{\Phi}(k)  }{\partial k_0}
+  \frac{\partial}{\partial k_j}\left[ \left( \frac{1-e^{-2\lambda k_0}}{2 \lambda}
- \frac{\lambda}{2} |\vec{k}|^2  \right) \tilde{\Phi}(k)\right] - \lambda k_j \tilde{\Phi}(k)  \right\} \delta(\tilde{k}^2)
& \nonumber
\\
&Q^{R}_j = \frac{i}{2} \int d^4 k  \, \frac{\tilde{k}_0}{\left| \tilde{k}_0\right|} \,
e^{ 2 \lambda k_0} \, \tilde{\Phi}(-k_0 ,-e^{\lambda k_0} \vec{k}) \epsilon_{jlm}
k_m \frac{\partial \tilde{\Phi}(k)}{\partial k_l}  \delta(\tilde{k}^2) ~.& \label{BothCharges}
\end{eqnarray}

In light of the findings reported in the previous section, we should stress that,
as the careful reader can easily verify, one can perform a Noether analysis
of pure space rotations ($\tau_j=0$ from the onset of the Noether analysis)
working with ordinary (commutative) transformation parameters $\sigma_j$, and obtain
exactly the charges $Q^{R}_j$. Therefore these charges can be meaningfully interpreted as
charges associated with the space-rotation symmetries of the theory.
Instead, since pure boosts are not allowed, one should perhaps be cautious in characterizing
the charges $Q^{N}_j$ as resulting from the invariance under boosts.

\section{Closing remarks}
The understanding of the physical significance of Hopf-algebra spacetime
symmetries must still be considered ``work in progress". The recently
acquired~\cite{k-Noether, freidKowa1, AntonMichSympl}
capability to bring to completion some Noether analyses should accelerate this progress,
but several corollary issues must be solved in order to fully exploit
this new tool.
Combining some findings reported in Ref.~\cite{k-Noether} and some of the observations
reported here, we are proposing a criterion for the search of a suitable description
of transformation parameters: these parameters should be allowed to be ``noncommutative"
(allowed
to be subject to nontrivial rules of commutation with the spacetime coordinates)
but should still act only by (associative)
multiplication on the spacetime coordinates.
While of course in this young subject it is appropriate to continue to probe the
robustness of all ideas, this criterion has passed already some nontrivial tests,
most notably leading to successful Noether analyses in all applications attempted so far.

Concerning the time-independent quantities, the charges, that our Noether analyses
produce a lot remains to be understood in order to establish whether
these time-independent quantities
really provide an acceptable formalization of quantities we measure in a laboratory.
The best way to illustrate our concerns is to consider the charge associated with
time-translation symmetry. How is that charge related to the observable measured
by, say, a calorimeter? The way in which a calorimeter works depends very strongly
on the law of conservation of energy-momentum in particle collisions, and this
is an aspect of $\kappa$-Minkowski theories that our Noether-analysis
techniques still do not allow us to master.
Some of these issues
could be addressed within a symmetry analysis of quantum fields in $\kappa$-Minkowski,
whereas here we only considered classical fields.

Still, the fact that within our line of analysis we encountered an obstruction for pure boosts
could be rather valuable. One of the primary motivations for
considering spacetime noncommutativity comes from the desire to develop
some intuition for the implications of non-classical, ``fuzzy", spacetimes,
but at least in the (not isolated) case of $\kappa$-Minkowski,
very little has been accomplished toward a physical characterization of the fuzzyness.
For example, attempts to describe the
observable ``distance between two $\kappa$-Minkowski spacetime points"
have shown very little promise.
One might perhaps, and the results reported here could provide a starting point for that,
attempt to characterize spacetime fuzzyness in terms of ``fuzzy limitations" for
symmetry transformations, rather than directly in terms of observables such
as distance, area and volume.

\bigskip
\bigskip
\bigskip

\vfil
\eject

$~~~~~~~~~~~~~~~~~~~~~~~~~~~~~~~~~~~~~~~~~~~~~~~~~~${\bf NOTE ADDED}

\medskip
While we were in the final stages of preparation of this manuscript, the paper in
Ref.~\cite{kowafreidnew} was posted on the arXiv, also reporting
an analysis of space-rotation and boost symmetries of some
theories in $\kappa$-Minkowski spacetime, but within a setup which is
completely different form the one we adopted here
and with interest directed toward different aspects of the theories.
The equation of motion (and the action) considered in
Ref.~\cite{kowafreidnew} are different from ours, and they are analyzed following
a different technical scheme. In particular,
Ref.~\cite{kowafreidnew} does not provide a formula for the charges carried by
classical fields in the noncommutative spacetime, but rather expresses the charges
in terms of some operators for the ``number of particles"
to be used in a corresponding theory of quantum fields.
Most importantly for us, Ref.~\cite{kowafreidnew} does not discuss the possibility
of limitations on the ``purity" of  Lorentz-sector transformations.
In Section III of Ref.~\cite{kowafreidnew} there is a discussion of Leibniz rule for
a Lorentz-sector differential transformation, but no comments are offered
on possible limitations to the types of pure transformation that are allowed.
And actually we find that (probably just as a result of a mere typographical
error) the equations reported in
Section III of Ref.~\cite{kowafreidnew} do not implement Leibniz rule for the
relevant differential transformation. The transformation parameters $\omega_{\mu \nu}$ introduced
in Section III of Ref.~\cite{kowafreidnew} would be ``noncommutative parameters"
of the type we advocate here, as implicitly codified in the equation
$e^{i \vec{k} \cdot \vec{x}} e^{- i k_0 x_0} \omega_{\mu \nu}
= \omega_{\alpha \beta} {K_{\mu \nu}}^{\alpha \beta} e^{i \vec{k} \cdot \vec{x}} e^{- i k_0 x_0}$,
but the properties later attributed to the matrix ${K_{\mu \nu}}^{\alpha \beta}$, which
are ${K_{\mu \nu}}^{\alpha \beta} = \delta_\mu^\alpha {k_\nu}^\beta$ with
${k_\nu}^\beta =
\left(
\begin{array}{c c}
1 & \vec{k} \\
0 & \mathbbm{1} e^{- k_0}
\end{array}
\right)$, would amount
to a ``no-pure-space-rotation" uncertainty principle
and would not comply with the Leibniz-rule requirement.

\end{document}